\documentclass[10pt,conference]{IEEEtran}

\IEEEoverridecommandlockouts
\ifCLASSOPTIONcompsoc
  \usepackage[nocompress]{cite}
\else
  \usepackage{cite}
\fi
\usepackage{amsmath,amssymb,amsfonts}
\usepackage{algorithmic}
\usepackage{graphicx}
\usepackage{textcomp}
\usepackage{xcolor}
\usepackage{url}
\usepackage{balance}
\usepackage[caption=false,font=normalsize,
   labelfont=sf,textfont=sf]{subfig}
\usepackage[nameinlink]{cleveref}
\usepackage{combelow} 
\usepackage{xspace}
\usepackage{multirow}
\usepackage{ascmac}
\usepackage{threeparttable}

\usepackage[utf8]{inputenc}
\usepackage[T1]{fontenc}
 \usepackage{arydshln}

\newcommand{\refAllARule}{\cite{TARMAQ,hyper_rule,age_and_length,practical_guideline,historank,morisan_journal,erose,ishida_ss,adaptive_asoc_rule,clone_and_arule}}

\begin{document}
\title{The Construction of an Empirical Dataset of\\ Incomplete Software Changes from\\ Open Source Projects}

\author{
    \IEEEauthorblockN{Savira Ramadhanty}
    \IEEEauthorblockA{
    \textit{Institute of Science Tokyo}\\
    Yokohama, Japan \\
    ORCID:0009-0005-8009-5648}
\and
    \IEEEauthorblockN{Profir-Petru P\^ar\cb{t}achi}
    \IEEEauthorblockA{
    \textit{Institute of Science Tokyo}\\
    Yokohama, Japan \\
    ORCID:0000-0003-4940-6864}
\and
    \IEEEauthorblockN{Yoshiya Ishida}
    \IEEEauthorblockA{
    \textit{Institute of Science Tokyo}\\
    Yokohama, Japan}
\and
    \IEEEauthorblockN{Takashi Kobayashi}
    \IEEEauthorblockA{
    \textit{Institute of Science Tokyo}\\
    Yokohama, Japan \\
    ORCID:0000-0002-2235-9992}
}

\maketitle

\begin{abstract}
During software development, a modification to a software component may propagate across the system, requiring precise identification and correct revision of all affected components. 
This is a complex task, and developers often (45.7\%) miss related changes.
To address this, several methods have been developed to extract co-change rules from files that are frequently changed together in the revision history. 
However, previous research evaluated the methods using artificially created incomplete changes, which may not be representative of real-world data. 
To solve this problem, we construct a dataset by mining incomplete changes from a collection of open-source software, using information about induced bugs and their respective fixes from an issue tracking platform. 
We also analyze the characteristics of incomplete changes using this constructed dataset and found that 89.4\% of missed changes involved five or fewer files. 
Finally, we re-evaluate LCExtractor, an existing co-change rule extraction method, on our constructed dataset, and we identify the optimal sorting criterion and the impact of the number of used commits.  
\end{abstract}


\section{Introduction}\label{sec:intro}
When a developer modifies a software component, that change can impact other components because of interdependencies.
Consequently, the developer must also modify the impacted components.
Otherwise, the change remains incomplete and can later manifest as bugs.
To prevent this, researchers have attempted to identify impacted components using static analysis~\cite{static_impact_analysis}.
However, static analysis cannot observe relationships among files whose static dependencies cannot be identified, such as build files.

To overcome the shortcomings of static analysis, previous work has applied data mining to revision histories, e.g., Git~\refAllARule{}.
These methods extract files that are frequently changed together in a revision history, identifying a relationship in which, when one file is changed, another file is likely to change as well.
They then provide developers with a list of files that should be changed together. In this paper, we call this type of method a co-change rule. 

There have been a number of methods based on the co-change rule method \cite{erose,TARMAQ,adaptive_asoc_rule}.
However, they are evaluated using artificially created incomplete changes.
In such evaluations, they selected one commit from the revision history and extracted the co-modified files. 
They then deliberately omitted some of the files.
The omitted files are treated as overlooked files, while the rest are treated as queries or files changed by developers.
The methods are evaluated based on whether they can recover the omitted files.
Such evaluations cannot determine the performance of the proposed methods on real-world incomplete changes. 

Furthermore, many prior studies evaluate change impact analysis based on individual developer modifications, such as those made in a single commit.
However, in branch-based development, changes may span multiple commits within a single branch.
In such cases, one possible approach is to perform change impact analysis by aggregating these multiple changes---for instance, when a pull request is created. 
Mori et al.~\cite{morisan_journal} have reported that grouping commits related to the same task can sometimes enable the extraction of high-quality association rules. 
Therefore, when evaluating change impact analysis, it is important to assess not only the analysis of individual changes but also the analysis of aggregated changes.
We treat this setting by exploring Multiple Inducing Commits as a setting which mimics the branch development setting instead of the squashed setting of Mori et al.

To address the artificial incomplete changes problem, in this research, we construct a dataset containing actual incomplete changes.
To do so, we focus on the relationship between induced bugs and their respective fixes, mined from issues recorded in an issue tracking system.
A portion of these issues are associated with the commit that fixes the bug (Bug Fixing Commit/BFC) and the commit that introduces that bug (Bug Inducing Commit/BIC).
We assume that when a bug caused by incomplete changes occurs, the BFC contains files that were not included in the BIC. 
Using the newly constructed dataset, we investigate the proportion of incomplete changes of all issue reports to understand their significance and characteristics in order to set a reference for co-change support design. 
Subsequently, we would also like to see how co-change rules perform in the constructed dataset, both in the case of a single change and aggregated multiple changes.

Once we construct our dataset, we address the following research questions.
First, to understand the significance of incomplete changes in software development, we investigate how often they occur in real-world environments (RQ1, Section \ref{sec:res:rq1}).
If incomplete changes constitute a considerable portion of committed changes, it highlights the need for robust change impact analysis tools. We found that incomplete changes make up of 45.6\% of software modifications.

Second, to design reliable tools that prevent incomplete changes, we must understand the nature of these omissions.
We need to identify how many files developers typically overlook and whether those files have historical co-change patterns.
By exploring these characteristics, we can determine what a recommendation tool must be capable of handling (RQ2, Section \ref{sec:res:rq2}).
After observing these characteristics, we discover that in incomplete changes, 49.5\% miss a single file and 89.4\% miss five or fewer. 
Additionally, co-change rules could have identified 69.7\% of these overlooked files.

Third, existing co-change methods rely on various association rule criteria to recommend files, but previous research investigate their optimal configuration using artificially created datasets.
Furthermore, since prior research mainly evaluated these criteria on single-commit changes, applying them to aggregated changes requires a thorough re-evaluation of these criteria.
Our analysis prove that Confidence is the optimal criterion for both single-commit and aggregated-commit changes (RQ3, Section \ref{sec:res:rq3}).

Finally, co-change rules are mined from revision histories.
As software evolves over time, its dependencies also change, which can make older historical data obsolete and introduce noise into the rule-mining process.
Therefore, we investigate whether older commit data affect the performance of co-change rules to determine if restrictions should be applied to the number of past commits used for mining (RQ4, Section \ref{sec:res:rq4}).
We found improved performance at higher mining ratio. 
This shows that old commits proves to be useful for identifying incomplete changes using co-change rules.

Overall, we make the following contributions:
\begin{itemize}
  \item We construct a dataset of actual incomplete changes by mining BFC-BIC pairs
  \item We analyze the characteristics of incomplete changes on the constructed dataset
  \item We reevaluate the performance of an existing co-change rule method on the constructed dataset
\end{itemize}

\section{Dataset Construction}\label{sec:dataset}
In this paper, we conduct analysis on file granularity, so the dataset we want to construct has to have the following properties:
\begin{itemize}
  \item A set of modified components (files) (later referred to as the query)
  \item A set of components (files) affected by the modification (later referred to as the expected outcome)
\end{itemize}
We construct the dataset in several steps.
First, we extract issues for each repository. Then, we keep the issues that track bug-fixing (Bug Fixing Issues/BFI) and their respective issues that caused the bugs (Bug Inducing Issues/BII).
After that, we find the commits related to the BFI and BII, identified the Bug Fixing Commit (BFC) and Bug Inducing Commit (BIC) respectively, and set them as pairs.
Finally, from these BFC-BIC pairs, we identify the overlooked files.
Figure \ref{fig:issue_commits} overviews this process, the green arrow represents the relationship used to form BFI-BII pairs, while the blue arrows allow us to translate these to BFC-BIC pairs.

The following subsections explain the steps in details.

\subsection{Issue crawling and filtering} \label{subsec:issue_crawling_filtering}

\begin{figure*}
  \centering
  \includegraphics[width=0.8\linewidth]{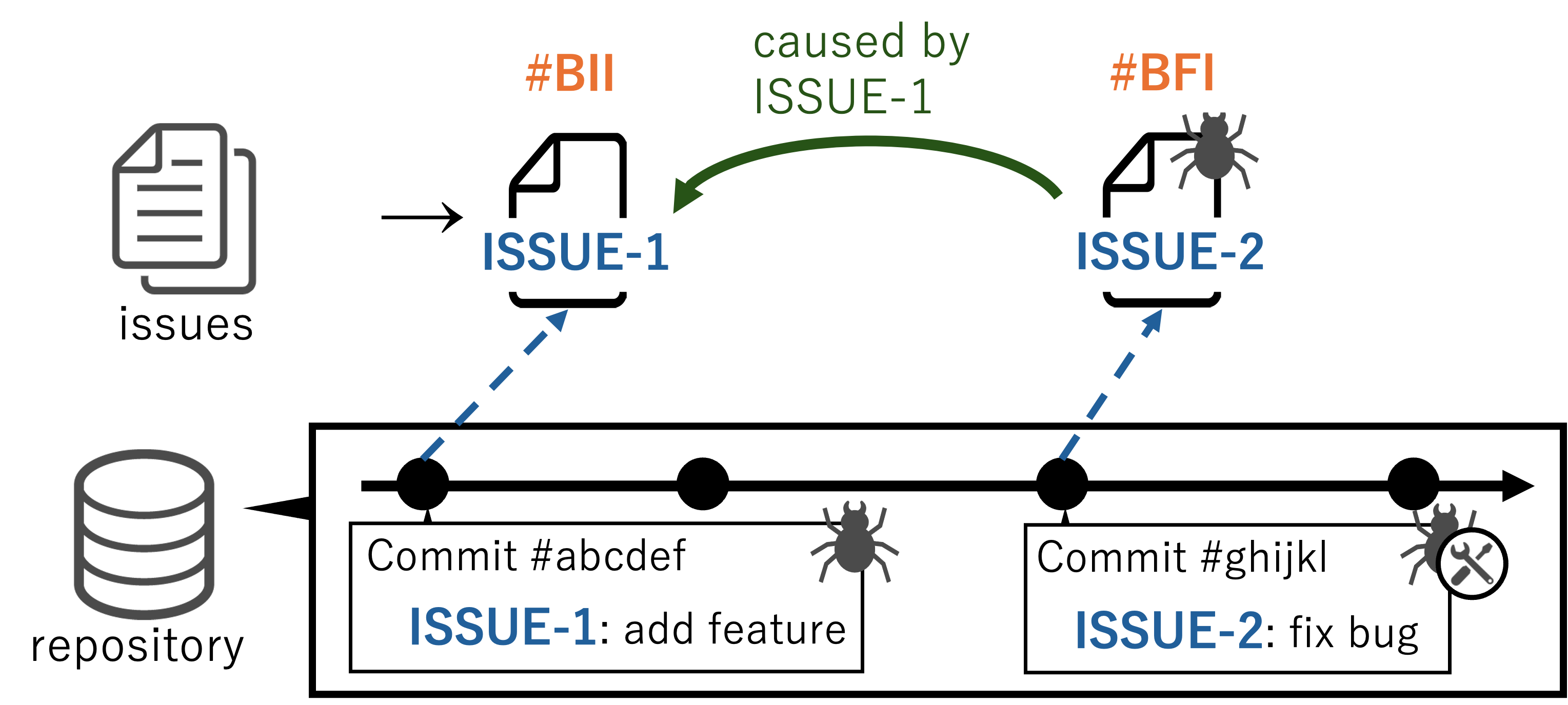}
  \caption{BFI-BII pair identification and ratio of issues. Commits in software repositories record the changes developers make. Issues are recorded separately to keep track of the reasons that necessitate the changes and are assigned a dedicated ID. To connect issues with associated changes, developers typically connect them by mentioning the issue ID in the commit message. Among issues, we use ``\textit{is broken by}'' and ``\textit{is caused by}'' links to identify bug-fixing issues (\#BFI) and bug-inducing issues (\#BII), and using their issue IDs, we trace the Bug-Fixing Commit (\#BFC) and Bug-Inducing Commit (\#BII), respectively.}
  \label{fig:issue_commits}
\end{figure*}

Well-documented software projects typically use an issue tracking system to manage problems and tasks throughout the project's life cycle.
In this research, we focus on Apache-owned projects, following Wen et al's method~\cite{induce_benchmark}.

Apache hosts its projects on GitHub and uses Jira as its issue-tracking system.
In Jira, each issue is assigned an ID, and issue links connect issues to establish relationships and dependencies between tasks.
To track bug-related tasks, Apache uses ``Bug'' as the issue type; when the bug is resolved, the bug report is closed. 
Wen et al. used ``\textit{is broken by}'' issue links to identify the causing-inducing issue pair~\cite{induce_benchmark}.
In addition, one of the projects uses ``\textit{is caused by}'' for the same purpose\footnote{\url{https://issues.apache.org/jira/browse/FLINK-14145}}.
Therefore, in this research, we use both link types to identify the issue that caused the bug.

We collect issues from the Apache projects used in Wen et al.'s method: Accumulo, Ambari, Hadoop, Lucene, and Oozie. 
We also collect issues from other projects that meet the following conditions, they have:
\begin{itemize}
  \item More than 500 GitHub stars;
  \item A Jira instance for that specific project;
  \item More than 30 issues fulfilling the following conditions:
  \begin{itemize}
    \item are resolved
    \item have a ``\textit{is broken by}'' or ``\textit{is caused by}'' link
  \end{itemize}
\end{itemize}
From that, we obtain the following additional projects: Calcite, Camel, Cassandra, Flink, HBase, Hive, Ignite, Lucene-Solr, Pig, Spark, Struts, Thrift, and Wicket.

We collect all issues from the specified projects. We then only focus on issues related to bugs. 
From this subset, we take the closed bug issues and identify the ones having ``\textit{is broken by}'' or ``\textit{is caused by}'' links. 
We call these Bug Fixing Issues (BFI), and we determine the issues pointed by these links as Bug Inducing Issues (BII). From these issues, we form BFI-BII pairs.

We use the BFI-BII pairs to construct our BFC-BIC dataset. Figure~\ref{fig:issue_commits} illustrates the process.
To relate specific actions in the software history to the issues, the issue ID is linked to the commit performing the action.
Previous work has shown that this traceability link is not perfectly recorded~\cite{missing_links};
however, the same work shows that more experienced developers are more likely to record this information.
Thus, if the problem we explore persists, it persists specifically for the same profile of developers.

\begin{figure}[t]
  \centering
  \includegraphics[width=\linewidth]{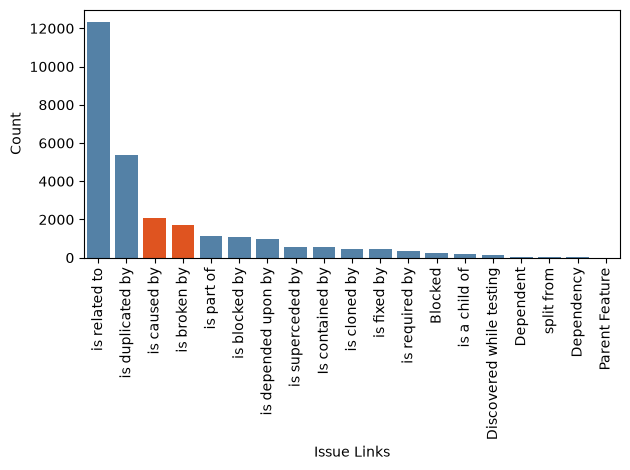}
  \caption{Distribution of issue links among the closed bug report issues in Apache projects we use, as described in Section~\ref{subsec:issue_crawling_filtering}}
  \label{fig:distribution_of_issue_links}
\end{figure}

Figure \ref{fig:distribution_of_issue_links} illustrates the distribution of issue links in the resolved bug report issues. 
The scarcity of ``\textit{is caused by}'' and ``\textit{is broken by}'' shows that the number of issues we can identify BFI from is very low.
\subsection{BFC-BIC pairs identification}\label{sec:dataset:bfc_bic}
We identify the corresponding BFCs and BICs by searching for commits that mention each issue ID in BFI-BII pairs.
However, there are cases where developers include issue ID in commit messages for purposes other than indicating change related to the issue.
To exclude such cases, we use a pattern to identify the issue ID within a commit.
For example, HBase recommends to include the issue ID at the beginning of the commit message\footnote{\url{https://github.com/apache/hbase/blob/master/src/main/asciidoc/_chapters/developer.adoc}}, and many commits follow such recommendation.
We use this pattern to link issues and the associated commits.
If a commit message mentions an Issue ID, we use only the issue ID mentioned in the pattern.
If all the commit messages do not include the pattern, we use all the issue ID to maximize issue-to-commit links~\cite{missing_links}.

We identify issue ID patterns from each project's contributor guide. 
For projects with no contributor guide, we determine the pattern by examining the commit messages manually.

\subsection{Identification of overlooked files}
Among BFC-BIC pairs, there are pairs having one BIC, and there are ones having multiple BICs.
We call the former Single Induce Data (SID) and the latter Multi Induce Data (MID).
For SIDs, we assume that change support is conducted when the BIC is created.
For MIDs, we consider a series of BICs as commits in a branch, that then are submitted as a pull request. We assume that changes in the branch will be combined.
Previous studies of change-support only evaluated on SIDs.
In our constructed dataset, with MIDs, it is possible to evaluate the change support method in the case where multiple changes are combined together.

To identify overlooked files, we extract the files modified in the BFC but not in the BIC.
However, it is also possible that BFC modifies files that are added after BIC.
Such cases are not overlooked files.
Therefore, when extracting file differences between BFCs and BICs, we made sure that the files had existed at the time when the BIC was committed.
Additionally, among the identified BICs, there are also commits with a creation date later than the BFC.
Because these commits do not contain the bug fixed by the BFC, we also exclude such commits from our dataset.

For the dataset collection, we use Apache projects described in Section \ref{subsec:issue_crawling_filtering}.
Since the projects have the same owner, they share the same issue documentation culture.

Table \ref{tab:used_project_apache} shows the list of projects we use in this research, along with the number of issues, identified BFI, BFI-BII pairs, and commits.

\begin{table}[t]
  \centering
  \footnotesize
  \caption{Apache projects used in this research}
  \label{tab:used_project_apache}
  {\normalsize
  \begin{tabular}{l | r : r r r}
    \textbf{Project} & \textbf{\# commits} & \textbf{\# issues} & \textbf{\# BFI} & \textbf{\# pairs} \\ \hline
    accumulo    & 14,968 & 4,745    & 2,213    & 120 \\
    ambari    & 25,090 & 26,300    & 16,927    & 47 \\
    calcite    & 6,632 & 7,621    & 3,443    & 83 \\
    camel    & 81,192 & 23,799    & 7,109    & 92 \\
    cassandra    & 31,891 & 21,247    & 10,221    & 248 \\
    flink    & 37,967 & 39,966    & 12,564    & 567 \\
    hadoop    & 28,261 & 17,606    & 7,178    & 351 \\
    hbase    & 21,136 & 30,027    & 12,173    & 310 \\
    hive    & 18,149 & 29,560    & 11,336    & 337 \\
    ignite    & 30,176 & 28,766    & 8,524    & 394 \\
    lucene-solr & 34,921 & 28,731    & 9,882    & 294 \\
    oozie    & 2,412 & 3,727    & 1,814    & 66 \\
    pig    & 3,767 & 5,397    & 2,613    & 63 \\
    spark    & 48,419 & 57,341    & 16,930    & 232 \\
    struts    & 8,293 & 5,568    & 3,070    & 64 \\
    thrift    & 7,706 & 6,069    & 2,852    & 44 \\
    wicket    & 22,231 & 7,178    & 4,211    & 118 \\
    \hline
    \textbf{Overall} & 423,211 & 343,648 & 133,060 & 3,430 \\
  \end{tabular}}
\end{table}

\begin{table}[t]
  \centering
  \caption{Interestingness Measures used in this research}
  \label{tab:used_im}
  {\normalsize
  \begin{tabular}{l | c}
    \textbf{Interestingness Measure}          &  \textbf{Definition}\\ \hline
    Confidence (conf) & $P(B|A)$                                   \\ 
    Support (supp) & $P(A,B)$                                   \\ 
    Lift (lift) & $\frac{P(A,B)}{P(A)P(B)}$                  \\ 
    Coverage (cov)   & $P(A)$                                     \\ 
    One Way Support (ows)   & $P(B|A) * \log_{2}\frac{P(A,B)}{P(A)P(B)}$  \\ 
    Prevalence (prev) & $P(B)$                                     \\ 
    Specificity (spec) & $P(\lnot B|\lnot A)$  \\ 
  \end{tabular}}
\end{table}

\section{Experiment}\label{sec:eval}

In this section, we first detail the RQs we wish to answer. We then detail the experimental setup for implementing co-change rules to identify overlooked files in the repositories from Section~\ref{subsec:issue_crawling_filtering}.
We then define the metrics used to evaluate their effectiveness and allowing us to answer our RQs.

\subsection{Research Questions}

In this research, we want to quantify how prevalent the overlooked change problem is, to characterise these incomplete changes, and to reassess previous results on our new dataset to see if performance or conclusions change.
To this end, we formulate the following research questions:

\textbf{RQ1: What is the percentage of incomplete changes out of all committed changes?}
In RQ1, we want to measure the prevalence of the incomplete changes problem.
To do so, we mine issues from the Apache projects we described in Section~\ref{subsec:issue_crawling_filtering}.
We then identify BFIs using ``\textit{is caused by}'' and ``\textit{is broken by}'' links, and keep only the ones having associated BIIs. These are the \textbf{\#pairs} described in Table~\ref{tab:used_project_apache}.
From a number of BFI-BII pairs, we obtain BIC-BFC pairs as described in Section~\ref{sec:dataset:bfc_bic}
From all the obtained BIC-BFC pairs, we calculate the percentage the number of pairs having overlooked changes over all the BFC-BIC pairs.

\textbf{RQ2: What are the characteristics of incomplete changes?}
We want to look at the nature of incomplete changes to set a reference for the development of change support methods.
To do so, for each of the identified incomplete changes in RQ1, we count the number of files involved.
Subsequently, we apply co-change rules on each project using the implementation described in Section~\ref{sec:eval:settings}.
We then calculate the proportion of files identifiable by co-change rules out of all overlooked files.

\textbf{RQ3: What is the most effective criterion for identifying incomplete changes?}
In RQ3, we want to determine the criterion used in the co-change rule that covers as many overlooked files as possible.
While previous work has a set standard of using Confidence and Support as the IMs, they were optimised on artificial data.
We wish to explore this setting on real-world data.
We do so by varying the Interestingness Measures in the co-change rule implementation as described in Section~\ref{sec:eval:settings}.

\textbf{RQ4: Do old commits introduce noise in the identification of incomplete changes?}
As file relationships in the early history can later become obsolete, in this RQ, we want to see whether old commits used to extract co-change rules compromise their effectiveness. 
We do so by varying the proportion of commits used in the co-change rule extraction, as described in Section \ref{sec:eval:settings}.

\subsection{Experiment Setting}\label{sec:eval:settings}

To evaluate the change-support method, one sorts co-change rules and measures the accuracy from the file rankings.
The Interestingness Measure (IM) is used as the criteria to sort the co-change rules.
There are about 40 kinds of IMs~\cite{hyper_rule,practical_guideline}.
Many previous studies~\cite{TARMAQ,erose,age_and_length,morisan_journal} use Confidence and Support, and they reported that using these measures resulted in high accuracy.
However, previous work used artificial data, so IMs may perform differently on our dataset.
Table \ref{tab:used_im} shows the IMs we compared in this study.
Additionally, researchers discussed about the number of commits used in extracting co-change rules~\cite{morisan_journal, age_and_length}.
As software development history becomes extensive, the dependencies between files can change, causing early history to become noise for later co-change information.

We evaluate the performance of co-change rules by varying the IMs in Table \ref{tab:used_im}.
All of them are definitions of the co-change rule A $\rightarrow$ B, which indicate that if file A is changed, then file B is also likely to be changed.
The IMs are expressed by the probability of occurrence in the commit that extracts the co-change rule. 
We also vary the number of commits from the constructed dataset used to extract co-change rules to verify its impact on the accuracy.

We use LCExtractor~\cite{morisan_journal} to extract the co-change rules. 
Additionally, when software branches are merged, CVS archives do not explicitly record the merge. 
Instead, all modifications from the branch appear as a single, massive change \cite{zimmermann2005mining}.
We wish to treat branch development through MIDs and such commits would introduce noise.
To prevent this, we exclude transactions that affect more than 30 files.
This helps keep the extracted rules focused on meaningful co-change relations.

\subsubsection{Number of used commits}
In this experiment, we vary the number of commits used to extract the co-change rule.
Previous studies use a predetermined number of commits, such as 5000 or 10,000~\cite{morisan_journal, TARMAQ,age_and_length}.
However, because the number of commits in the dataset varies, we use a ratio instead of exact numbers.
This ratio is the proportion of commits used for co-change rule extraction out of all commits in each project.
In this paper, we call this the mining ratio.
We use four mining ratios: 1.0, 0.75, 0.5, and 0.25.

\subsubsection{Co-change rule sorting criteria}
We sort co-change rules by combining three types of IMs.
This is because combining two IMs may not always be sufficient to resolve ties~\cite{ishida_ss}.
Co-change rules that remain tied even after combining three IMs are sorted based on the file path names of the recommended files.
We refer to this IM as the "sorting criterion." 

When combining three Interestingness Measures, simply selecting three out of the seven available options produces inappropriate combinations.
For instance, if we choose Support as the primary criterion and Coverage as the secondary, we cannot effectively use Confidence as the tertiary criterion.
Because association rules with identical Support and Coverage share the exact same underlying terms ($P(A, B)$ and $P(A)$), their Confidence values will also be identical. 
Consequently, using Confidence as a third criterion to break ties serves no purpose.
Therefore, we conduct our experiments by eliminating any combinations where the third measure provides no meaningful distinction.

\subsubsection{Evaluation Metrics}

We use 6 metrics to evaluate the performance of co-change rules: Average Precision (AP), Mean Average Precision (MAP), Top 20 Rank, Recall@20, and Recall. The definitions of each metric are as follows:
\begin{itemize}
  \item \textbf{Average Precision (AP)}
  
  For a given dataset, let $recom$ be the ranked list of recommended files; the value is calculated using:
  \begin{eqnarray}
    AP = \sum_{k=1}^{|recom|} p(k) * \Delta r(k)
    \label{eq:ap_calc}
  \end{eqnarray}
  $p(k)$ represents the precision when the top $k$ items in $recom$ are recommended.
  $\Delta r(k)$ denotes the difference in recall between the top $k$ recommendations and the top $k-1$ recommendations. 
  \item \textbf{Mean Average Precision (MAP)}

  MAP is the mean of the AP values across the entire dataset. We use this to determine the overall accuracy.
  \item \textbf{Top 20 Rank}
  
  The number of relevant files found in the top 20 ranks.
  Top 20 Rank represents the proportion of data for which at least one file was correctly recommended within the top 20 positions.
  Letting $D$ be the set of data, $recom_{i}^{20}$ be the set of recommended files (up to the 20th rank) for a given data item, and $TP_{i}$ be the set of files that were overlooked during modification, Top 20 Rank is calculated using Equation \ref{eq:success_rate_at_k}.
  \begin{eqnarray}
    Top\:20\:Rank = \frac{1}{|D|} \sum_{i=1}^{|D|} \begin{cases}
                                                      1 & (|recom_{i}^{20} \cap TP_{i}| > 0) \\
                                                      0 & (otherwise)
                                                    \end{cases}
    \label{eq:success_rate_at_k}
  \end{eqnarray}
  \item \textbf{Recall@20}
  
  Recall@20 evaluates the comprehensiveness of recommendations in the top 20 recommended items. 
  For each data point, recall@20i is calculated using Equation \ref{eq:recall_k}, and Recall@20 is determined by averaging these values.
  \begin{eqnarray}
    recall@20_{i} = \frac{|recom_{i}^{20} \cap TP_{i}|}{|TP_{i}|}
    \label{eq:recall_k}
  \end{eqnarray}
  \item \textbf{Recall}
  
  Recall evaluates the comprehensiveness of recommendations without the constraints imposed by recommendation rankings.
\end{itemize}
When we evaluate the co-change rules, we apply it to the query, and we evaluate whether the expected outcome can be recommended, or recommended with a high rank.
We identify the overlooked files from BFC-BIC pairs, as well as the query and expected outcome from them. 
Consider a case where a bug occurs in a software due to a change for bug-fixing or feature addition.
If the cause of the bug is that the developer did not change a file that should be changed at the same time, then the BFC will change the file that was overlooked in the BIC.
Based on this, we consider files modified in BFC but not in BIC to be the files developers forget to change in the BIC. We consider this as the expected outcome.
We set the files modified in BIC as the query.

\section{Results}
In this section, we first explore the prevalence of overlooked changes (RQ1).
We then investigate the number of files involved in those overlooked changes and whether co-change rules can identify them (RQ2).
Subsequently, we identify the most effective criterion used in co-change rules (RQ3).
Finally, we analyze the impact of the number of commits used to extract co-change rules (RQ4).
\subsection{\textbf{RQ1: Percentage of incomplete changes}}\label{sec:res:rq1}

\begin{figure}[t]
  \centering
  \includegraphics[width=0.5\linewidth]{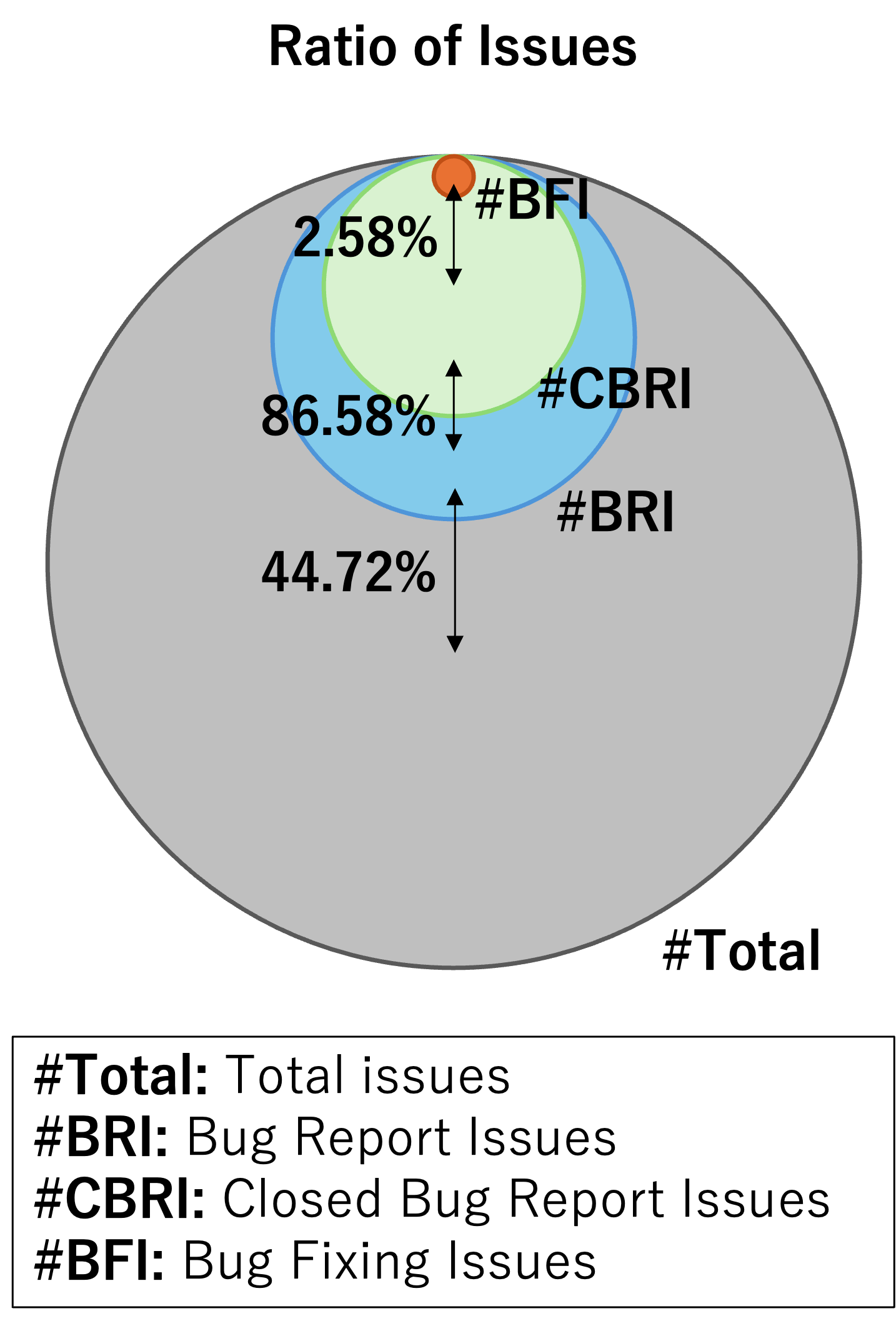}
  \caption{The ratio of BFI over the total issues for all Apache projects described in Section~\ref{subsec:issue_crawling_filtering}}
  \label{fig:ratio_issues_total}
\end{figure}
Following the procedures that Section \ref{sec:dataset} outline, we have created our overlooked changes dataset by mining data from the selected projects from the earliest issues to issues created in July 2026.
Figure \ref{fig:ratio_issues_total} illustrates the BFIs over the total issues in the Apache projects mentioned in Section~\ref{subsec:issue_crawling_filtering}.
Of all collected issues, 44.72\% are Bug issues (Bug Report Issues/\#BRI), 
86.58\% of \#BRI are resolved (Closed Bug Report Issues/\#CBRI), 
and only 2.58\% of \#CBRI are linked to Bug Inducing Issues with the specified link types (Bug Fixing Issues/\#BFI), 
showing that even in well-documented projects like Apache projects, documented BFI-BII pairs are rare.
We then lifted the BFI-BII pairs into BFC-BFI pairs.

In total, we obtain 2,428 bug reports linked to fix commits and inducing issues, and 789 of them contain incomplete changes.
The percentage of incomplete changes in the constructed dataset is 45.7\%.
This shows that, even in well-documented software projects, nearly half of the changes committed for bug-fixing are incomplete.
Table \ref{tab:incomplete_changes_percentage} shows the percentage of incomplete changes in it.
We obtained 789 SIDs and 320 MIDs.
\begin{table}[t]
  \centering
  \begin{threeparttable}
  \caption{Percentage of incomplete changes in our dataset}
  \label{tab:incomplete_changes_percentage}
  {\normalsize
  \begin{tabular}{l | c c c c }
    \textbf{Project} & \textbf{\#pairs} & \textbf{\#SID} & \textbf{\#MID} & \textbf{\#Incomp} \\
    \hline
    accumulo    & 101 & 12  & 29  & 0.406 \\
    ambari      & 38  & 10  & 4   & 0.368 \\
    calcite     & 64  & 23  & 5   & 0.438 \\
    camel       & 61  & 13  & 11  & 0.393 \\
    cassandra   & 189 & 72  & 17  & 0.471 \\
    flink       & 363 & 86  & 76  & 0.446 \\
    hadoop      & 266 & 73  & 17  & 0.338 \\
    hbase       & 232 & 91  & 20  & 0.478 \\
    hive        & 290 & 132 & 16  & 0.510 \\
    ignite      & 183 & 83  & 10  & 0.508 \\
    lucene-solr & 158 & 38  & 47  & 0.538 \\
    oozie       & 61  & 23  & 4   & 0.443 \\
    pig         & 60  & 30  & 8   & 0.633 \\
    spark       & 188 & 66  & 14  & 0.426 \\
    struts      & 47  & 11  & 17  & 0.596 \\
    thrift      & 36  & 9   & 5   & 0.389 \\
    wicket      & 91  & 17  & 20  & 0.407 \\
    \hline
    \textbf{Overall}& \textbf{2,428} & \textbf{789} & \textbf{320} & \textbf{0.457} \\
  \end{tabular}}
  \begin{tablenotes}[flushleft]\footnotesize
    \item \textbf{\#pairs}: the number of identified BIC-BFC pairs
    \item \textbf{\#SID}: the number of SID
    \item \textbf{\#MID}: the number of MID
    \item \textbf{\#Incomp}: the percentage of BFC-BIC pairs that actually
contain incomplete changes
  \end{tablenotes}
  \end{threeparttable}
\end{table}

\begin{screen}
  \textbf{Answer to RQ1:} Incomplete changes make up 45.7\% of the bug-fixing changes.
\end{screen}

\subsection{\textbf{RQ2: Characteristics of incomplete changes}}\label{sec:res:rq2}

To investigate the characteristics of incomplete changes, we first analyze the number of files involved in each overlooked change. 
Additionally, we determine the proportion of these missed files that co-change rules could identify.
Table \ref{tab:incomplete_changes_characteristics} shows the statistical results of the characteristics of incomplete changes.
We find that the percentage of data having one overlooked file is 40\%, except for Ignite, with an overall percentage of one overlooked file of 49.5\%.
On the other hand, the percentage of data having equal to or fewer than 5 overlooked files in the dataset is always above 80\%,
This shows that when incomplete changes actually occur, most of the overlooked files are still equal to or fewer than 5.

$RuleCoverage$ shows the proportion of files identifiable by co-change rules among the specified overlooked files.
We obtain this by using all the commits available in each project.
In our dataset, co-change rules can identify between 36.0\% to 92.2\% of the overlooked files.
On average, they could identify 63.9\% overlooked files.
From this, we infer that co-change rules are still effective on actual incomplete changes.
Additionally, these findings suggest the importance of detecting incomplete changes and confirm the usefulness of co-change rules for this purpose.
\begin{screen}
  \textbf{Answer to RQ2:} When incomplete changes occur, 49.5\% involved an overlooked change in only a single file, while 89.4\% involved five or fewer overlooked files. Additionally, 69.7\% of the overlooked files could have been identified by co-change rules.
\end{screen}
\begin{table}[t]
  \centering
  \begin{threeparttable}
  \caption{Characteristics of incomplete changes in our dataset}
  \label{tab:incomplete_changes_characteristics}
  {\normalsize
  \begin{tabular}{l | c c c }
    \textbf{Project} & \textbf{ratio of 1} & \textbf{ratio of $\leq$ 5} & \textbf{RuleCoverage} \\
    \hline
    accumulo	& 0.634	& 0.927	& 0.588 \\
    ambari	& 0.429	& 0.857	& 0.629 \\
    calcite	& 0.500	& 0.929	& 0.867 \\
    camel	& 0.458	& 0.958	& 0.833 \\
    cassandra	& 0.494	& 0.910	& 0.841 \\
    flink	& 0.457	& 0.833	& 0.540 \\
    hadoop	& 0.611	& 0.978	& 0.720 \\
    hbase & 0.559 & 0.874 & 0.729 \\
    hive & 0.473 & 0.892 & 0.666 \\
    ignite	& 0.366	& 0.860	& 0.731 \\
    lucene-solr	& 0.494	& 0.894	& 0.881 \\
    oozie & 0.519	& 0.926	& 0.883 \\
    pig & 0.553	& 0.947 & 0.922 \\
    spark & 0.475	& 0.900	& 0.800 \\
    struts & 0.571	& 0.893	& 0.360 \\
    thrift	& 0.429	& 0.929	& 0.615 \\
    wicket	& 0.429	& 0.929	& 0.500 \\
    \hline
    \textbf{Overall}	& \textbf{0.495}	& \textbf{0.894}	& \textbf{0.697} \\
  \end{tabular}}
  \begin{tablenotes}[flushleft]\footnotesize
    \item \textbf{ratio of 1}: the proportions of data points where overlooked files are 1
    \item \textbf{ratio of $\leq$ 5}: the proportions of data points where overlooked files are 5 or fewer
    \item \textbf{RuleCoverage}: the proportion of files identifiable using co-change rules among the specified overlooked files
  \end{tablenotes}
  \end{threeparttable}
\end{table}

\subsection{\textbf{RQ3: Most effective criterion for identifying incomplete changes}}\label{sec:res:rq3}

We use a score for each criterion to identify the optimal one for identifying overlooked files.
We consider a sorting criterion that frequently achieves a high MAP to be the optimal criterion.
To calculate the score, we use Equation \ref{eq:sort_criteria_score}.
\begin{eqnarray}
  score = \sum_{i=1}^{3} \frac{1}{i} * kth\_highest\_time(i)
  \label{eq:sort_criteria_score}
\end{eqnarray}
For each mining ratio, we identify the sorting criterion that achieved the $i$-th highest MAP (where $1 \leq i \leq 3$). 
$kth\_highest\_time(i)$ represents the number of times the criterion achieved the $i$-th highest MAP.
We use four mining ratios; the maximum possible score is 4.

Table \ref{tab:single_score_result} shows the calculated sorting criterion scores for SIDs, and Table \ref{tab:multi_score_result} shows them for MIDs. 
The "1st," "2nd," and "3rd" columns indicate the criteria used as the primary, secondary, and tertiary criteria, respectively. 
Where criteria are separated by a slash ("/"), the resulting score is the same regardless of which one is used. 
In both SID and MID, the sorting criteria with the highest score used Confidence as the primary criterion.
This indicates Confidence is the optimal sorting criterion. 
\begin{table}[t]
  \begin{center}
    \caption{Scoring criteria used on SID}
    \label{tab:single_score_result}
    {\normalsize
    \begin{tabular}{l l l | r}
      \textbf{1st} & \textbf{2nd} & \textbf{3rd} & \textbf{Score} \\ \hline
      conf	&  cov/supp	& prev	& 4.000 \\
      conf	&  cov/supp	& lift/ows/spec	& 1.500 \\
      conf	&  prev	& cov/spec/supp	& 1.167 \\
      supp	&  conf	& lift/ows/prev/spec	& 0.333 \\
    \end{tabular}}
  \end{center}
\end{table}

\begin{table}[t]
  \begin{center}
    \caption{Scoring criteria used on MID}
    \label{tab:multi_score_result}
    {\small
    \begin{tabular}{l l l | r}
      \textbf{1st} & \textbf{2nd} & \textbf{3rd} & \textbf{Score} \\ \hline
      conf	&  prev	& cov/spec/supp	& 4.000 \\
      supp	&  conf/cov/lift	& lift/ows/spec/conf/prev/cov	& 0.500 \\
      supp	&  conf/cov/lift/ows	& prev/cov/spec/conf/lift	& 0.333 \\
    \end{tabular}}
  \end{center}
\end{table}

\begin{screen}
  \textbf{Answer to RQ3:} The optimal sorting criterion for both SID and MID is Confidence as the primary criterion. 
\end{screen}

\subsection{\textbf{RQ4: Impact of old commits on co-change information}}\label{sec:res:rq4}

We examine how old commits affect co-change rules in identifying incomplete changes by adjusting the mining ratio across four different values. 
Tables \ref{tab:single_score_result_mining_ratio} and \ref{tab:multi_score_result_mining_ratio} present the values of the evaluation metrics for SID and MID, respectively. 
For each mining ratio, we measured the MAP, Top 20 Rank, Recall@20, and Recall. 

In both SID and MID, the maximum metrics are achieved when the mining ratio is higher (1.0 and 0.75). 
Conversely, it decreases when the mining ratio is lower (0.5 and 0.25).
This shows that old commits actually provide useful information for co-change rules instead of introducing noise.

We observe the effectiveness of co-change rules as a change support method against actual incomplete changes based on Tables \ref{tab:single_score_result_mining_ratio} and \ref{tab:multi_score_result_mining_ratio}.
Looking at Table \ref{tab:single_score_result_mining_ratio}, in SID, the MAP value is low, and Recall@20 is 35-40\% of the total recall.
This shows that overlooked files cannot be identified within the top 20 recommendations. 
The highest Top 20 Rank metric is around 38\%, which is lower than the reported Top 20 Rank metric in the initial dataset, which is around 40\%. 
This means roughly one or two out of five cases, at least one overlooked file is identified within the top 20 results.
Nevertheless, the co-change rules demonstrated applicability to actual incomplete changes to recommend overlooked changes within a certain range.

\begin{screen}
  \textbf{Answer to RQ4:} Higher performance on higher mining ratios shows that old commits provide useful information for identifying incomplete changes using co-change rules.
\end{screen}

\begin{table}[t]
  \centering
  \caption{Performance on SID with different mining ratios}
  \label{tab:single_score_result_mining_ratio}
  {\normalsize
  \begin{tabular}{l | r r r r}
    \multirow{2}{*}{\textbf{Metrics}} & \multicolumn{4}{c}{\textbf{Mining Ratio}} \\
    & 1.0 & 0.75 & 0.5 & 0.25 \\ 
    \hline
    MAP & \textbf{0.110}  & \textbf{0.110}  & 0.106 & \underline{0.103} \\
    Top 20 Rank & 0.386 & \textbf{0.387} & 0.385 & \underline{0.361} \\
    Recall @ 20 & 0.268 & 0.268 & \textbf{0.270}  & \underline{0.253} \\
    Recall  & \textbf{0.764} & 0.748 & 0.725 & \underline{0.631} \\
  \end{tabular}}
\end{table}

\begin{table}[t]
  \centering
  \begin{threeparttable}
  \caption{Performance on MID with different mining ratios}
  \label{tab:multi_score_result_mining_ratio}
  {\normalsize
  \begin{tabular}{l | r r r r}
    \multirow{2}{*}{\textbf{Metrics}} & \multicolumn{4}{c}{\textbf{Mining Ratio}} \\
    & 1.0 & 0.75 & 0.5 & 0.25 \\ 
    \hline
    MAP	& \textbf{0.095}	& 0.094	& 0.093	& \underline{0.091} \\
    Top 20 Rank	& 0.291	& \textbf{0.297}	& \underline{0.286}	& 0.287 \\
    Recall @ 20	& 0.183	& \textbf{0.186}	& 0.178	& \underline{0.176} \\
    Recall	& \textbf{0.771}	& 0.767	& 0.705	& \underline{0.610} \\
  \end{tabular}}
  \begin{tablenotes}[flushleft]\footnotesize
    \item \textbf{Bold} values indicate the maximum
    \item \underline{Underlined} values indicate the minimum
  \end{tablenotes}
\end{threeparttable}
\end{table}

\section{Discussion}

The high prevalence of incomplete changes, which occur in nearly half of bug-fixing changes, uncovers a vulnerability in software development process.
This frequency highlights the necessity for change-support methods to prevent these omissions from turning into subsequent bugs.
The fact that co-change rules could capture nearly 70\% of these omissions demonstrate the usefulness of historical data.

In Section~\ref{sec:res:rq2}, we observe that these omissions typically involve five or fewer files, indicating that developers rarely miss massive changes, but rather are overlooking closely related files.
Because these files are closely related in the historical data, co-change rules can identify them.
However, the remaining missing files indicate that information about files changing together is not effective enough, and we can explore other information in the historical data to identify more overlooked files in the future.

When evaluating the extraction of these rules, we observe that Confidence is the optimal criterion for both SID and MID.
This implies that regardless of the number of changes committed, the conditional probability of files changing together provides more reliable information rather than just mere frequency.
Comparing the SID and MID results from Tables \ref{tab:single_score_result_mining_ratio} and \ref{tab:multi_score_result_mining_ratio} reveals that MID yields higher overall recall, whereas SID yields higher MAP, Top 20 Rank, and Recall@20 values.
This is because MID aggregates multiple changes, making it easier to extract a larger number of co-change rules.
While this allows for the identification of overlooked files, it also results in the extraction of false positives, thereby degrading the rank-related metrics. 
Despite this decline, MID still achieves a Top 20 Rank of approximately 28\%.
This implies that a missed change can be identified within the top 20 results roughly once every three attempts, which is still practical. 

\section{Threats to Validity}\label{sec:threats}

Our dataset construction hinges on two categories of links: two types of issue-issue links and commit-issue links.
There is a risk that either or both of these categories of links is not recorded; for the latter, previous work documents this issue~\cite{missing_links}.
However, the bias that this introduces, based on observations from the same Bachmann et al.~\cite{missing_links} paper, is what we would call a ``good developer'' bias.
That is, the developer that is more likely to be observed by our data collection by virtue of recording these metadata links, is more likely to have higher experience.
In our setting, if the issue of overlooked changes persists even for such experienced developers, it is a sign that the real rate is likely even higher and we observe a lower estimate.

Further, we focus on Apache projects. This does bias our results to the project culture and development practices from Apache.
Previous work has similarly focused on Apache due to their mature development processes and higher quality data.

Finally, the number of commits in each project used to extract co-change rules varied significantly; therefore, we determined the number of used commits based on specific ratios.  
However, the accuracy of different change support methods may exhibit different results on different projects despite using the same ratio of used commits.

\section{Related Works}\label{sec:relworks}

This section discusses previous research relevant to change support methods.
First, we explore existing change-support methods that use revision history, with a specific focus on those utilizing co-change rules.
Second, because our method relies on information from bug-fixing commits and their respective bug-inducing commits,
we discuss previous studies on how to identify bug-inducing commits.

\subsection{Change support using revision history}
Early change support methods applied association rules to the revision history.
Zimmerman et al. proposed ROSE~\cite{erose}, which extracted co-change rules from revision history using association rules and used them to suggest software entities developers should change.
They conducted experiments to determine whether ROSE could successfully recommend entities that developers had overlooked when committing changes to a version control system, demonstrating the accuracy of ROSE's recommendations. 
However, ROSE uses strict rules to extract the rules, resulting in few situations where recommendations can be made to developers.

After ROSE, several studies proposed methods for extracting co-change rules to uncover more dependencies that ROSE could not identify to its strictness.
Rolfsnes et al. proposed TARMAQ~\cite{TARMAQ}, which used a more relaxed requirement than ROSE for extraction. 
They demonstrated that TARMAQ is fast and outperforms ROSE in terms of Average Precision
They also proposed Hyper-Rule, which aggregates several co-change rules~\cite{hyper_rule}. 
Pugh et al. proposed adaptive targeted association rule mining~\cite{adaptive_asoc_rule} to extract co-change rules from fewer commits. 
Hagward developed LCExtractor~\cite{hagward2015using-obsolute}, which can extract more rules without applying constraints.

In another research direction, Moonen et al. investigated interestingness measures (of which over 40 exist across various domains~\cite{geng_interestingness_measure}), as well as the number of changed files that limit the number of commits used in co-change rules~\cite{practical_guideline}.
They also investigated how the number of commits used in co-change rule extraction (history length) and the number of commits added after co-change rule extraction affect the performance of change support~\cite{age_and_length}. 
Mondal et al. combine information about co-change rules and code clones~\cite{clone_and_arule}.
Mondal et al. also compared 5 types of co-change rule sorting and proposed HistoRank, which dynamically changes the sorting method based on the accuracy of past change support~\cite{historank}.
Mori et al. considered temporal proximity and integrated commits associated with the same task to extract co-change rules~\cite{morisan_journal}. 
Ishida et al. also investigated the effectiveness of Hyper-Rule on co-change rules extracted by LCExtractor and the effectiveness of setting the second criterion for sorting co-change rules~\cite{ishida_ss}.
All the methods in previous studies above were evaluated on artificially created incomplete changes. 
This biases reported performance which may differ in real-world scenarios.

\subsection{BIC identification}
{\'S}liwerski used the SZZ algorithm to identify BIC~\cite{szz_first}.
Following this, later works introduced improvements using the annotation graph~\cite{szz_improve_2006} and line mapping algorithm~\cite{szz_revisited}. 
On the other hand, the accuracy of BIC identification has been contested, and studies report that there are a certain number of cases where the BIC is incorrectly identified~\cite{induce_benchmark}.

Other research explored ways to identify BIC. 
As described in Section \ref{subsec:issue_crawling_filtering}, Wen et al. use issues from Apache projects that contain information about the cause of bugs to identify BFC and BIC to make the InduceBenchmark dataset~\cite{induce_benchmark}. 
Additionally, some of the BFC-BIC pairs identified in InduceBenchmark involve multiple BICs. 
However, relying on the pairs in InduceBenchmark is insufficient to have an adequate volume of data, particularly for cases involving multiple BICs.

\section{Conclusion}\label{sec:concl}
Evaluations of change support methods based on co-change rules previously relied on artificially created incomplete changes derived from commits; however, this approach cannot determine the performance of the methods on actual incomplete changes.
To address this limitation, we construct a dataset of incomplete changes by analyzing the relationship between Bug-Fixing Commits (BFCs) and Bug-Introducing Commits (BICs) and investigated the characteristics of incomplete changes.

Our analysis reveals that incomplete changes occurred in approximately half of the identified BFC-BIC pairs, and these incomplete changes predominantly involved a small number of overlooked files. 
When applying co-change rules, the results confirm that "Confidence" is the effective sorting metric for co-change rules. 
Additionally, we find that older commits provide useful information for change support instead of introducing noise.

As future directions, we plan on using the constructed dataset to evaluate more change-support methods besides co-change rules.
Because our dataset explicitly links bug-fixing and bug-inducing commits, this enables the analysis of other software development tasks that utilize this information.
Additionally, having observed that co-change rules still have room for improvement, we plan to use more information from revision history beyond co-change patterns to further enhance the change-support tool.

\vspace{0.3em}
\noindent\textbf{Acknowledgement: }
During the preparation of this work the authors used ChatGPT and Gemini in order to improve readability and language of the work. 
After using this tool/service, the authors reviewed and edited the content as needed and take full responsibility for the content of the publication.

\bibliographystyle{IEEEtran-tklab}
\bibliography{IEEEabrv,ref}

@inproceedings{static_impact_analysis,
  title={Using coupling measurement for impact analysis in object-oriented systems},
  author={Briand, L C. and others},
  xbooktitle={Proc. ICSM},
  booktitle={Proceedings IEEE International Conference on Software Maintenance (ICSM). },
  pages={475--482},
  year={1999}
}

@inproceedings{TARMAQ,
  title={Generalizing the analysis of evolutionary coupling for software change impact analysis},
  author={Rolfsnes, T and others},
  xbooktitle={Proc. SANER},
  booktitle={2016 IEEE 23rd International Conference on Software Analysis, Evolution, and Reengineering (SANER)},
  pages={201--212},
  year={2016}
}

@inproceedings{hyper_rule,
  title={Improving change recommendation using aggregated association rules},
  author={Rolfsnes, T and others},
  xbooktitle={Proc. MSR},
  booktitle={2016 IEEE/ACM 13th Working Conference on Mining Software Repositories (MSR)},
  pages={73--84},
  year={2016}
}

@inproceedings{age_and_length,
  title={Exploring the effects of history length and age on mining software change impact},
  author={Moonen, L and others},
  xbooktitle={Proc SCAM},
  booktitle={2016 IEEE 16th International Working Conference on Source Code Analysis and Manipulation (SCAM)},
  pages={207--216},
  year={2016}
}

@inproceedings{practical_guideline,
  title={Practical guidelines for change recommendation using association rule mining},
  author={Moonen, L and others},
  xbooktitle={Proc. ASE},
  booktitle={Proceedings of the 31st IEEE/ACM International Conference on Automated Software Engineering (ASE)},
  pages={732--743},
  year={2016}
}

@inproceedings{historank,
  title={HistoRank: History-Based Ranking of Co-change Candidates},
  author={Mondal, M and others},
  xbooktitle={Proc. SANER},
  booktitle={2020 IEEE 27th International Conference on Software Analysis, Evolution and Reengineering (SANER)},
  pages={240--250},
  year={2020}
}

@article{morisan_journal,
  jatitle	 = {改版履歴の分析に基づく変更支援手法における時間的近接性の考慮と同一作業コミットの統合による影響},
  title   = {An Empirical Study of the Effects of Recency-aware History Analysis and Commits Aggregation on Change Guide},
  author	 = {Mori, T and others},
  jajournal = {情報処理学会論文誌},
  journal	 = {IPSJ Journal},
  volume	 = {58},
  number	 = {4},
  pages	 = {807--817},
  year 	 = {2017},
}

@article{erose,
  title={Mining version histories to guide software changes},
  author={Zimmermann, T and others},
  journal={IEEE Transactions on Software Engineering},
  volume={31},
  number={6},
  pages={429--445},
  year={2005},
}

@TechReport{ishida_ss,
  jaauthor          = {石田 and 小林},
  author          = {Y. Ishida and T. Kobayashi},
  jatitle          = {改版履歴分析に基づく変更漏れ防止支援における変更ルール集約と順位付けの効果},
  title           = {Effects of rule aggregation and ranking method on change history analysis based error prevention methods},
  xinstitution =  {IEICE},
  type = 	 {IEICE Technical Report},
  year = 	 {2019},
  number = 	 {SS2018-65},
  month = 	 {Mar.},
  OPTnote = 	 {(Vol.118, No.471, pp.79--84)},
  OPTannote = 	 {}
}

@inproceedings{adaptive_asoc_rule,
  title={The case for adaptive change recommendation},
  author={Pugh, S and others},
  xbooktitle={Proc. SCAM},
  booktitle={2018 IEEE 18th International Working Conference on Source Code Analysis and Manipulation (SCAM)},
  pages={129--138},
  year={2018},
  xorganization={IEEE}
}

@inproceedings{clone_and_arule,
  title={Associating Code Clones with Association Rules for Change Impact Analysis},
  author={Mondal, M and others},
  xbooktitle={Proc. SANER},
  booktitle={2020 IEEE 27th International Conference on Software Analysis, Evolution and Reengineering (SANER)},
  pages={93--103},
  year={2020}
}

@Misc{hagward2015using-obsolute,
  title={Using Git Commit History for Change Prediction: An empirical study on the predictive potential of file-level logical coupling},
  author={Hagward, A},
  year={2015}
}

@inproceedings{szz_first,
  title={When do changes induce fixes?},
  author={{\'S}liwerski, J and others},
  booktitle={Proc. International Workshop on Mining Software Repositories (MSR)},
  pages={1--5},
  year={2005},
  xjournal={ACM sigsoft software engineering notes},
  xvolume={30},
  xnumber={4},
}

@InProceedings{szz_improve_2006,
  title={Automatic identification of bug-introducing changes},
  author={Kim, S and others},
  xbooktitle={Proc. ASE},
  booktitle={21st IEEE/ACM International Conference on Automated Software Engineering (ASE)},
  pages={81--90},
  year={2006},
  xorganization={IEEE},
  OPTmonth = 	 {},
  OPTnote = 	 {},
  OPTannote = 	 {}
}

@inproceedings{szz_revisited,
  title={Szz revisited: verifying when changes induce fixes},
  author={Williams, C and Spacco, J},
  xbooktitle={Proc. DEFECTS},
  booktitle={Proceedings of the 2008 Workshop on Defects in Large Software Systems (DEFECTS)},
  pages={32--36},
  year={2008}
}

@inproceedings{induce_benchmark,
  title={Exploring and exploiting the correlations between bug-inducing and bug-fixing commits},
  author={Wen, M and others},
  xbooktitle={Proc. ESEC/FSE},
  booktitle={Proceedings of the 2019 27th ACM Joint Meeting on European Software Engineering Conference and Symposium on the Foundations of Software Engineering (ESEC/FSE))},
  pages={326--337},
  year={2019}
}

@article{geng_interestingness_measure,
author = {Geng, Liqiang and Hamilton, Howard J.},
title = {Interestingness measures for data mining: A survey},
year = {2006},
issue_date = {2006},
xpublisher = {Association for Computing Machinery},
xaddress = {New York, NY, USA},
volume = {38},
number = {3},
issn = {0360-0300},
xurl = {https://doi.org/10.1145/1132960.1132963},
xdoi = {10.1145/1132960.1132963},
xjournal = {ACM Comput. Surv.},
journal = {ACM Computing Survey},
month = sep,
pages = {9–es},
numpages = {32}
}

@inproceedings{missing_links,
author = {Bachmann, Adrian and Bird, Christian and Rahman, Foyzur and Devanbu, Premkumar and Bernstein, Abraham},
title = {The missing links: bugs and bug-fix commits},
year = {2010},
isbn = {9781605587912},
xpublisher = {Association for Computing Machinery},
xaddress = {New York, NY, USA},
xurl = {https://doi.org/10.1145/1882291.1882308},
doi = {10.1145/1882291.1882308},
booktitle = {Proceedings of the Eighteenth ACM SIGSOFT International Symposium on Foundations of Software Engineering (FSE)},
xbooktitle = {Proc. FSE},
pages = {97–106},
numpages = {10},
location = {Santa Fe, New Mexico, USA},
series = {FSE '10}
}

@article{zimmermann2005mining,
  title={Mining version histories to guide software changes},
  author={Zimmermann, Thomas and Zeller, Andreas and Weissgerber, Peter and Diehl, Stephan},
  journal={IEEE Transactions on software engineering},
  volume={31},
  number={6},
  pages={429--445},
  year={2005},
  publisher={IEEE}
}

\end{document}